\documentclass[conference,10pt,compsocconf]{IEEEtran}
\usepackage{cite}
\usepackage{amsmath,amssymb,amsfonts}
\usepackage{algorithmic}
\usepackage{graphicx}
\usepackage{textcomp}
\usepackage{xcolor}
\usepackage{authblk}
\usepackage{hyperref}

\usepackage{amsthm}

\usepackage{booktabs}

\renewcommand{\v}[1]{{\bf #1}}
\newcommand{\m}[1]{{\bf #1}}
\newcommand{\R}{\mathbb{R}}
\newcommand{\E}[1]{\langle #1 \rangle}

\renewcommand{\O}{\mathcal{O}}
\DeclareMathOperator{\tr}{Tr}

\DeclareFixedFont{\ttb}{T1}{txtt}{bx}{n}{11} 
\DeclareFixedFont{\ttm}{T1}{txtt}{m}{n}{11}  
\DeclareFixedFont{\ttbs}{T1}{txtt}{bx}{n}{9} 

\definecolor{deepblue}{rgb}{0,0,0.5}
\definecolor{deepred}{rgb}{0.6,0,0}
\definecolor{deepgreen}{rgb}{0,0.5,0}

\usepackage{listings}

\newcommand\pythonstyle{\lstset{
language=Python,
basicstyle=\ttm,
otherkeywords={self},             
keywordstyle=\ttb\color{deepblue},
commentstyle=\ttm\color{gray},
emph={MyClass,__init__},          
emphstyle=\ttb\color{deepred},    
stringstyle=\color{deepgreen},
frame=tb,                         
showstringspaces=false            %
}}

\lstnewenvironment{python}[1][]
{
\pythonstyle
\lstset{#1}
}
{}

\usepackage{algorithm}
\usepackage{algorithmic}

\hypersetup{colorlinks=true,urlcolor=blue,linkcolor=.,citecolor=.}
\def\BibTeX{{\rm B\kern-.05em{\sc i\kern-.025em b}\kern-.08em
    T\kern-.1667em\lower.7ex\hbox{E}\kern-.125emX}}
\begin{document}

\title{Efficient Principal Subspace Projection of Streaming Data Through Fast Similarity Matching
}

\author[*1]{Andrea Giovannucci}
\author[*1]{Victor Minden}
\author[1]{Cengiz Pehlevan}
\author[1,2]{Dmitri B. Chklovskii}

\affil[1]{Center for Computational Biology, Flatiron Institute, New York, NY 10010 \authorcr Email: {\tt \{agiovannucci, vminden, cpehlevan, mitya\}@flatironinstitute.org}\vspace{1.5ex}}
\affil[2]{Neuroscience Institute, NYU Langone Medical Center, New York, NY 10016  \vspace{-2ex}}

\maketitle

\begin{abstract}
Big data problems frequently require processing datasets in a streaming fashion, either because all data are available at once but collectively are larger than available memory or because the data intrinsically arrive one data point at a time and must be processed online. Here, we introduce a computationally efficient version of similarity matching \cite{derivation}, a framework for online dimensionality reduction that incrementally estimates the top $K$-dimensional principal subspace of streamed data while keeping in memory only the last sample and the current iterate. To assess the performance of our approach, we construct and make public a test suite containing both a synthetic data generator and the infrastructure to test online dimensionality reduction algorithms on real datasets, as well as performant implementations of our algorithm and competing algorithms with similar aims. Among the algorithms considered we find our approach to be competitive, performing among the best on both synthetic and real data.
\end{abstract}



\section{Introduction}

Reducing the dimensionality of large, high-dimensional datasets is often a crucial step in data analysis. Perhaps the most popular and widely used technique is to project the data to a $K$-dimensional principal subspace, in the sense of Principal Component Analysis (PCA) \cite{pearson1901liii}. Here we introduce an efficient algorithm for this principal subspace projection (PSP) task that operates in the very commonly used online setting.

Often a dataset is not available in entirety, but received in a streaming fashion, one datum at a time. Also often is the case when a dataset is larger than the available memory and needs to be processed in chunks. For both cases, it is imperative to develop PSP algorithms that operate online and process data one at a time. Because of its importance, this problem has attracted a lot of attention in the literature, see e.g.  \cite{sanger1989optimal,foldiak1989adaptive,rubner1990development,kung1994adaptive,yang1995projection,diamantaras1996principal,warmuth2008randomized,arora,goes2014robust,crammer2006online,derivation}. In this paper, we focus on a setting where the algorithm needs to output a new estimate of the principal subspace after every new datum, i.e., mini-batch is not allowed. We assume limited memory so that the algorithm can only store $\O(DK)$ real numbers going from one datum to another, where $D$ is the number of dimensions that the data lives in. We remark that this is the minimum order attainable, because the $K$ principal components would have a total of $DK$ elements.

A recently introduced promising algorithm that operates in the setting described above is the Similarity Matching (SM) algorithm \cite{derivation}. Here, we introduce an efficient modification of the SM algorithm for online PSP, called Fast Similarity Matching (FSM), that reduces the cost per iteration from $\O(DK +K^3)$ to $\O(DK)$, i.e., the cost of reading the current iterate. We numerically demonstrate that with this modification, the FSM algorithm is competitive and often better performing than state-of-the-art algorithms, both in operation time and in convergence rate. It is not our intention here to do a full comparison of a large number of existing algorithms for online PCA/PSP, see instead Cardot and Degras \cite{cardot} for an excellent survey. Rather, we choose the top performing algorithms from Cardot and Degras for benchmarking, namely the Incremental PCA (IPCA) \cite{arora} and the Candid, Covariance-free IPCA (CCIPCA) algorithms \cite{ccipca},  and show that FSM is competitive by performing detailed numerical tests on synthetic and real datasets. As a second major contribution, we provide a package for online PCA/PSP with performant implementations in Python and MATLAB of the SM, FSM, IPCA, and CCIPCA algorithms.

The rest of the paper is organized as follows. In Section II, we introduce some notation. In Section III, we review the SM, IPCA and CCIPA algorithms and introduce the FSM algorithm. In Section IV, we describe our numerical simulations and software. In Section V, we present our numerical results. We conclude in Section VI.

\section{Notation}
We assume in our discussion that the data are stationary with mean $\E{\v{x}_t} = \v{0}\in\R^D$ and population covariance matrix $\E{\v{x}_t\v{x}_t^\top} = \m{C}\in\R^{D\times D}$.  In the batch setting, we assume we have $N$ data points $\{\v{x}_t\}_{t=1}^N$ with which we can define the batch covariance matrix $\m{C}_\text{batch} = \frac{1}{N}\sum_{t=1}^N \v{x}_t\v{x}_t^\top$.  With $\m{C}=\m{U}\m{\Sigma^2}\m{U}^\top$ as an eigendecomposition, we write the top principal components of $\m{C}$ as $\m{U}_K\in\R^{D\times K}$ (which we note are defined only up to a sign ambiguity).  Analogously, we write the principal components of the batch covariance matrix as $\m{U}^\text{(batch)}_K$, though this is primarily of interest for our real data examples.  We use hats to denote estimated quantities.

\section{Algorithms}\label{sec:alg}
In what follows, we begin by reviewing the similarity matching algorithm and then introduce a modification to improve asymptotic complexity.  After that, we review two competing algorithms that have been previously demonstrated \cite{cardot} to offer competitive performance in terms of accuracy or runtime.

\subsection{Similarity Matching (SM)}\label{sec:sm}
Previously, Pehlevan and collaborators introduced \cite{derivation} and analyzed \cite{neural_computation} the SM framework for online PSP, using a multidimensional scaling objective function to derive a neural network architecture. Starting with the batch optimization problem
\begin{align}\label{eq:sm}
\min_{\{\v{y}_t\}_{t=1}^N\subset \R^K} \sum_{s=1}^N\sum_{t=1}^N (\v{x}_s^\top\v{x}_t - \v{y}_s^\top\v{y}_t)^2,
\end{align}
the solution of which is given by projecting each $\v{x}_t$ onto the $K$-dimensional principal subspace $\v{U}_K$, the SM framework converts the minimization problem \eqref{eq:sm} to a saddle-point formulation amenable to online optimization as follows.   First, expanding the square and dropping the constant term gives
\begin{align}\label{eq:sm2}
\min_{\{\v{y}_t\}_{t=1}^N} \sum_{s=1}^N\sum_{t=1}^N -2\v{x}_s^\top\v{x}_t\v{y}_t^\top\v{y}_s + \v{y}_s^\top\v{y}_t\v{y}_t^\top\v{y}_s.
\end{align}  Then, we may introduce dummy optimization variables $\m{W}~\in~\R^{K\times D}$ and $\m{M}~\in~\R^{K\times K}$ to obtain
\begin{align*}
\min_{\{\v{y}_t\}_{t=1}^N} \min_{\m{W}} \max_{\m{M}}
 \quad &2\tr(\m{W}^\top\m{W}) - \tr(\m{M}^\top\m{M})\\
 - &4 \sum_{t=1}^N \v{x}_t^\top \m{W}^\top\v{y}_t + 2\sum_{t=1}^N \v{y}_t^\top \m{M}^\top\v{y}_t ,
\end{align*}
where optimizing over $\m{M}$ and $\m{W}$ and plugging in the optimal solutions recovers \eqref{eq:sm2}.  Finally, using duality to perform the optimization over $\{\v{y}_t\}_{t=1}^N$ explicitly gives the SM saddle-point problem
\begin{align}\label{eq:sm-saddle}
\min_\m{W} \max_\m{M} \quad 2\tr(\m{W}^\top\m{W}) - \tr(\m{M}^\top\m{M}) - 2 \sum_{t=1}^N \v{x}_t^\top \m{W}^\top\v{y}_t,
\end{align}
where $\v{y}_t \equiv \m{M}^{-1}\m{W}\v{x}_t$ for $t=1,2,\dots,N$.

Using simultaneous stochastic gradient steps for $\m{M}$ and $\m{W}$ in \eqref{eq:sm-saddle} gives Algorithm \ref{alg:sm}, where the rows of $\m{M}^{-1}\m{W}$ at convergence are orthogonal vectors spanning the principal subspace.  This illustrates a distinction between SM and the other algorithms we consider (and, in general, between PSP and PCA): SM in this form can recover the span of the top singular vectors but not the singular vectors themselves.  In other words, we can obtain an approximation
\begin{align}\label{eq:smrecon}
\m{\widehat U}_{K}\m{\widehat U}_{K}^\top = \m{W}^\top\m{M}^{-2}\m{W},
\end{align}
but we cannot completely recover an estimate of $\m{U}_K$ due to a rotational degeneracy, i.e., we have
$$
\m{W}^\top\m{M}^{-1} = \m{\widehat U}_K\m{Q},
$$
where $\m{\widehat U}_K$ is an estimate of the top principal components but $\m{Q}$ is an unknown rotation..  For the problem of determining a linear subspace capturing maximal variance of the data, however, this is not an important consideration.

We remark that using simultaneous stochastic gradient descent/ascent on \eqref{eq:sm-saddle} is not provably convergent in general, but is a popular scheme for online optimization of saddle point problems that can be effective in practice.

\begin{algorithm}
 \caption{Similarity Matching (SM) \cite{derivation,neural_computation}}
 \begin{algorithmic}[1]\label{alg:sm}
 \renewcommand{\algorithmicrequire}{\textbf{Input:}}
 \REQUIRE Initial weights $\m{M}\in\R^{K\times K}$ and $\m{W}\in\R^{K\times D}$
  \FOR {$t = 1,2,3,\dots$}
  \STATE\texttt{// Project x\_t into current space}
  \STATE $\v{y}_t \gets \m{M}^{-1}\m{W} \v{x}_t$
  \STATE\texttt{// Update weights}
  \STATE $\m{W} \gets (1-\alpha_t)\m{W} + \alpha_t \v{y}_t\v{x}_t^\top$
  \STATE $\m{M} \gets (1-\beta_t)\m{M} + \beta_t \v{y}_t\v{y}_t^\top$
  \ENDFOR
 \end{algorithmic}
 \end{algorithm}

 \subsection{An efficient modification: Fast Similarity Matching (FSM)}

 The complexity of each iteration of SM in Algorithm \ref{alg:sm} is $\O(DK + K^3)$, where the $\O(K^3)$ cost stems from solving the linear system
 $$
 \m{M}\v{y}_t = \m{W}\v{x}_t
 $$
 to resolve Line 7.  For problems where $K$ is large (e.g., $K\sim \sqrt{D}$ or greater), the cost of the linear solve is non-negligible and dominates the asymptotic cost of an iteration, which is both a theoretical and practical concern.  This motivates us to consider a modified approach.

 As can be seen in Algorithm \ref{alg:sm}, at each iteration we make a rank-one update to $\m{M}$ on Line 6, which we must subsequently use to solve a linear system on Line 3 at the next iteration. To exploit this fact, we can use the well-known Sherman-Morrison formula for updating the inverse of a matrix in response to a rank-one change (see, e.g., Hager \cite{hager}), which gives
 \begin{align*}
 \left(\m{A} + \v{v}\v{v}^\top\right)^{-1} = \m{A}^{-1} - \frac{\m{A}^{-1}\v{v}\v{v}^\top\v{A}^{-1}}{1+\v{v}^\top\m{A}^{-1}\v{v}},
 \end{align*}
 for $\m{A}\in\R^{M\times M}$ and $\v{v}\in\R^M$.
 Applying this to our case, we see that if
 $$
 \m{M}^+ =(1-\beta_t) \m{M} + \beta_t \v{y}_t\v{y}_t^\top
 $$
 then with $c_t \equiv \frac{\beta_t}{1-\beta_t}$ we have
 \begin{equation}\label{eq:minvformula}
 \begin{aligned}
 \left( \m{M}^+\right)^{-1} &= \frac{1}{1-\beta_t}\left(\m{M} + c_t \v{y}_t\v{y}_t^\top\right)^{-1}\\
 &= \frac{1}{1-\beta_t}\left(\m{M}^{-1} -\frac{c_t\m{M}^{-1}\v{y}_t\v{y}_t^\top\m{M}^{-1}}{1+c_t\v{y}_t^\top\m{M}^{-1}\v{y}_t} \right),
 \end{aligned}
 \end{equation}
 which shows that if $\m{M}^{-1}$ is already known then $\left( \m{M}^+\right)^{-1}$ can be found efficiently with complexity only $\O(K^2)$.

 Therefore, to increase efficiency we re-parameterize Algorithm \ref{alg:sm} in terms of the inverse of $\m{M}$ instead of $\m{M}$ itself, giving our modified Fast Similarity Matching (FSM) algorithm,  Algorithm \ref{alg:fsm}.  Numerically, the results of Algorithm \ref{alg:sm} and Algorithm \ref{alg:fsm} are the same in infinite precision.  However, compared to SM we see that the complexity of an iteration of FSM has now dropped to $\O(DK)$ as the linear solve has been eliminated.  With this modification, the per-iteration complexity of FSM is the same as reading the current iterate, which is a lower bound for any algorithm that updates the entire iterate at each step.

 \begin{algorithm}
 \caption{Fast Similarity Matching (FSM)}
 \begin{algorithmic}[1]\label{alg:fsm}
 \renewcommand{\algorithmicrequire}{\textbf{Input:}}
 \REQUIRE Initial weights $\m{M}_\text{inv}\in\R^{K\times K}$ and $\m{W}\in\R^{K\times D}$
  \FOR {$t = 1,2,3,\dots$}
  \STATE\texttt{// Project x\_t into current space}
  \STATE $\v{y}_t \gets \m{M}_\text{inv}\m{W} \v{x}_t$
  \STATE\texttt{// Update weights using \eqref{eq:minvformula}}
  \STATE $\m{W} \gets (1-\alpha_t)\m{W} + \alpha_t \v{y}_t\v{x}_t^\top$
  \STATE $\m{M}_\text{inv} \gets \frac{1}{1-\beta_t}\m{M}_\text{inv}$
  \STATE $\v{z}_t \gets \m{M}_\text{inv} \v{y}_t$
  \STATE $\m{M}_\text{inv} \gets \m{M}_\text{inv} - \frac{\beta_t}{1+\beta_t\v{z}_t^\top\v{y}_t} \v{z}_t\v{z}_t^\top$
  \ENDFOR
 \end{algorithmic}
 \end{algorithm}

\subsection{Other algorithms}

For comparison, we consider two other algorithms that we describe in the remainder of the section.

\subsubsection{Incremental PCA (IPCA)}

The IPCA algorithm of Arora et al.\ \cite{arora} is obtained by adapting the incremental singular value decomposition of Brand \cite{brand,brand2} to approximate singular vectors of the covariance matrix $\m{C}$ (see also Bunch et al.\ \cite{bunch}).

At iteration $t$, suppose that we have a rank-$K$ estimate $\m{\widehat C}_{K}$ of the covariance matrix $\m{C}$, given by
$$
\m{\widehat C}_{K} \equiv \m{\widehat U}_{K}\m{\widehat\Sigma}_{K}^2\m{\widehat{U}}_{K}^\top.
$$
The algorithm of Arora et al.\ computes the factors of the updated estimate $\m{\widehat C}^+_{K}$ by incorporating the new data point $\v{x}_t$ and then projecting the result back to a rank-$K$ matrix, i.e.,
$$
\m{\widehat C}^+_{K} = \mathcal{P}_{K}\left(\m{\widehat U}_{K}\m{\widehat\Sigma}_{K}^2\m{\widehat{U}}_{K}^\top + \alpha_t \v{x}_t\v{x}_t^\top\right),
$$
where $\mathcal{P}_{K}(\m{A})$ is the closest rank-$K$ matrix to $\m{A}$ in Frobenius norm and we typically take $\alpha_t = 1/t$.
Of course, for computational efficiency it is necessary to avoid ever forming the full matrix $\m{\widehat C}^+_K$, which is accomplished by following the steps in Algorithm \ref{alg:ipca}.

As noted in Arora et al., it is possible to construct examples where IPCA does not converge, and as such a general proof of convergence does not exist.  However, in practice we observe no convergence issues.
\begin{algorithm}
 \caption{Incremental PCA (IPCA) \cite{arora}}
 \begin{algorithmic}[1]\label{alg:ipca}
 \renewcommand{\algorithmicrequire}{\textbf{Input:}}
 \REQUIRE Initial squared singular value estimates $\m{\widehat\Sigma}_{K}^2\in\R^{K\times K}$ and singular vector estimates  $\m{\widehat U}_{K}\in\R^{D\times K}$
  \FOR {$t = 1,2,3,\dots$}
  \STATE\texttt{// Project x\_t into current space}
  \STATE $\v{y}_t \gets \m{\widehat U}^\top_{K}\v{x}_t$
  \STATE\texttt{// Compute residual of x\_t}
  \STATE $\v{r}_t \gets \v{x}_t - \m{\widehat U}_{K}\v{y}_t$
    \STATE \texttt{// Compute top K eigenpairs of a (K+1)-by-(K+1) matrix}\vspace{0.2cm}
  \STATE $\m{M} \gets (1-\alpha_t) \left[\begin{array}{cc}\m{\widehat\Sigma}_{K}^2& \v{0}\\\v{0}&0\end{array} \right] + \alpha_t \left[\begin{array}{cc}\v{y}_t\v{\tilde x}^\top& \|\v{r}_t\| \v{y}_t^\top\\ \|\v{r}_t\| \v{y}_t &\|\v{r}_t\|^2\end{array} \right]$\vspace{0.2cm}
 \STATE $(\m{V}_K,\,\m{\Lambda}_K) \gets \text{eig}(\m{M},K)$\vspace{0.2cm}
  \STATE $\m{\widehat\Sigma}_{K}^2 \gets \m{\Lambda}_K$\vspace{0.2cm}
  \STATE $\m{\widehat U}_{K} \gets  \left[\begin{array}{cc}\m{\widehat U}_{K}& \frac{\v{r}_t}{\|\v{r}_t\|}\end{array} \right]\m{V}_K$\vspace{0.2cm}
  \ENDFOR
 \end{algorithmic}
 \end{algorithm}

The asymptotic complexity of a single iteration of IPCA in Algorithm \ref{alg:ipca} is the highest among the three algorithms we consider, with a cost of $\O(DK^2 + K^3)$ due to the matrix-matrix multiplication on Line 10 and the eigendecomposition on Line 8.

\subsubsection{Candid, Covariance-free IPCA (CCIPCA)}
The CCIPCA algorithm of Weng et al.\ \cite{ccipca} effectively combines the standard stochastic power method (or normalized Hebbian rule, see, e.g., the discussion in Oja \cite{ojaold}) for a single component $\v{\hat u}$,
$$
\v{\hat u}^+ = \frac{(1-\alpha_t) \v{\hat u} + \alpha_t \v{x}_t\v{x}_t^\top\v{\hat u} }{\left\|(1-\alpha_t) \v{\hat u} +\alpha_t \v{x}_t\v{x}_t^\top\v{\hat u}\right\|},
$$
with a deflation scheme.  This is illustrated in Algorithm \ref{alg:ccipca}.

 The argument for convergence of principal components in CCIPCA is intuitive.  Because the estimates for $k=1$ are running stochastic power iteration (see Algorithm \ref{alg:ccipca}), $\v{\hat u}_1$ and $\hat\sigma_1^2$ eventually converge (for stationary input).  At this point, the estimates for $k=2$ look like stochastic power iteration on a modified data stream, with the modified data being orthogonal to $\v{\hat u}_1$ (Line 9).  Therefore, $\v{\hat u}_2$ and $\hat\sigma_2^2$ gradually converge, and so on.  A formal analysis can be found in Zhang and Weng \cite{zhang}.
\begin{algorithm}
 \caption{Candid, Covariance-free IPCA (CCIPCA) \cite{ccipca}}
 \begin{algorithmic}[1]\label{alg:ccipca}
 \renewcommand{\algorithmicrequire}{\textbf{Input:}}
\REQUIRE Initial squared singular value estimates $\hat\sigma_k^2$ and singular vector estimates $\v{\hat u}_{k}\in\R^D$ for $k=1,2,\dots,K$
  \FOR {$t = 1,2,3,\dots$}
  \STATE $\v{x} \gets \v{x}_t$
  \FOR {$k=1,2,\dots,K$}
    \STATE\texttt{// Compute new estimate of k-th singular pair}
    \STATE $\v{v} \gets (1-\alpha_t) \hat\sigma_k^2\v{\hat u}_k + \alpha_t \v{x}\v{x}^\top \v{\hat u}_k$
    \STATE $\hat\sigma_k^2 \gets \|\v{v}\|$
    \STATE $\v{\hat u}_k \gets \v{v} / \|\v{v}\|$
    \STATE\texttt{// Deflate}
    \STATE $\v{x} \gets \v{x} -  \v{\hat u}_k\v{\hat u}_k^\top \v{x}$
  \ENDFOR
  \ENDFOR
 \end{algorithmic}
 \end{algorithm}

 The computational complexity of an outer iteration of CCIPCA in Algorithm \ref{alg:ccipca} is $\O(DK)$, as each inner iteration can be seen to involve $\O(D)$ work.  From an implementation standpoint, it is worth noting that, due to deflation, the inner iterations must be performed sequentially.  Unlike the other algorithms we consider CCIPCA does not directly compute the projection of $\v{x}_t$ onto the current estimated subspace, but this can easily be incorporated at minimal additional cost.

\section{Numerical simulations and software}\label{sec:numres}
To evaluate the performance of the algorithms in Section \ref{sec:alg} in a systematic way, we implement each algorithm in a common software framework allowing for standardized tests with both synthetically generated data and real-world test data.  In this section, we describe the data and framework for numerical simulations.

The corresponding Python software package {\ttbs{online\_psp}} is available on GitHub\footnote{\normalsize\url{https://github.com/flatironinstitute/online_psp}} and can be used as a performant implementation of the described algorithms, as well as a vehicle for reproducing the results in Section \ref{sec:results}.  A simple example of the interface can be seen in the code snippet below, where we instantiate a class for PSP using FSM, iterate over the data points, and then recover a matrix whose columns span the estimate of the principal subspace.

\begin{python}
fsm_class = FSM(K, D)
for x in X:
    fsm_class.fit_next(x)
# Recover Uhat up to rotation Q
UhatQ = fsm_class.get_components()
\end{python}
An analogous MATLAB package with a subset of the functionality, {\ttbs{online\_psp\_matlab}}, is also available.\footnote{\normalsize\url{https://github.com/flatironinstitute/online_psp_matlab}}

\subsection{Synthetic data: the spiked covariance model}\label{sec:syn}
A popular model for synthetic data that we use here is the spiked covariance model \cite{johnstone}, which effectively generates multivariate Gaussian observations constrained to a low-dimensional subspace and then adds a small amount of isotropic Gaussian noise.

Formally, the model we use is a zero-mean multivariate Gaussian $\m{x} \sim \mathcal{N}\left(\v{0},\m{C}\right)$
with covariance matrix
\begin{align}\label{eq:truecov}
\m{C} = \E{\v{x}\v{x}^\top}= \m{U}_*\m{S}\m{U}_*^\top + \rho\m{I},
\end{align}
where $\m{U}_*\in\R^{D\times K}$ has orthonormal columns, $\m{S}\in\R_+^{K\times K}$ is diagonal and $\rho\in\R_+$ is a parameter dictating the noise level.

For the population principal components $\m{U}_*$, we  construct an orthonormal basis for a random $K$-dimensional subspace of $\R^D$ by first sampling $\m{Z}\in\R^{D\times K}$ with independent standard normal entries $Z_{ij} \sim \mathcal{N}(0,1)$ and then computing a thin QR factorization $\m{Z} = \m{Q}\m{R},$ where $\m{Q}\in\R^{D\times K}$ has orthonormal columns and $\m{R}\in\R^{K\times K}$ is upper triangular.  Then, we can take $\m{U}_*=\m{Q}$.

To sample from the model, assuming $\v{z}_t\in\R^{K}$ and $\v{w}_t\in\R^{D}$ are vectors with independent standard normal components, we can compute
\begin{align}\label{eq:sample}
\v{x}_t = \m{U}_*\m{S}^{1/2}\v{z}_t +  \sqrt{\rho}\v{w}_t,
\end{align}
from which it is readily verified that $\m{x}_t$ has the desired moments.

\subsection{Real data}
For our simulations on real-world data sets, we choose three popular datasets with varying dimensionality $D$ and number of examples $N$:  YALE\cite{mat_faces}, ORL\cite{ATT}, and MNIST \cite{lecun1998gradient}.

The ORL\cite{ATT} face database contains ten different images for 40 distinct subjects acquired under different lighting conditions, facial expressions and facial details (with/without glasses). The YALE dataset, as downloaded from a public database \cite{mat_faces}, includes 64 near frontal images of 38 individuals  under different lighting conditions. MNIST\cite{lecun1998gradient}  is a popular handwritten digits database, comprising 60,000 examples size-normalized and centered in a fixed-size image. These are summarized in Table \ref{tab:realdata}.

\begin{table}
\centering
  \caption{Details of the real datasets used in our simulations\label{tab:realdata}}
  \setlength{\tabcolsep}{0.35em}
\begin{tabular}{@{}lccl@{}}
\toprule
Name& $N$ & $D$ & Description  \\ \midrule
  YALE\cite{mat_faces} & 2432 &$1024$ ($32\times 32$)  & faces     \\
  ORL   \cite{ATT} & 400 & $11224$ ($112\times 92$) & faces  \\
  MNIST\cite{lecun1998gradient} &   60000& $784$ ($28\times 28)$ & digits   \\\bottomrule
\end{tabular}
\end{table}

\subsection{Performance metric: accuracy}\label{sec:accuracy}
To assess the statistical accuracy of the algorithms considered we choose a metric that we refer to as the \emph{subspace error} used previously by Cardot and Degras \cite{cardot}.  Given a matrix $\m{U}\in\R^{D\times K}$ with orthonormal columns, the orthogonal projector onto the range of $\m{U}$ is $\m{P}_\m{U}\equiv\m{U}\m{U}^\top$.  Crucially, $\m{P}_\m{U}$ depends only on the subspace spanned by columns of $\m{U}$ and is invariant to rotations of $\m{U}$ such as discussed in Section \ref{sec:sm}.

Given two matrices $\m{U}\in\R^{D\times K}$ and $\m{V}\in\R^{D\times K}$ each with orthogonal columns, we can therefore measure the difference between the range of $\m{U}$ and the range of $\m{V}$ in a basis-independent way via
\begin{align}\label{eq:err}
\text{Err}(\m{U},\m{V}) &\equiv \frac{\left\| \m{P}_\m{U} - \m{P}_\m{V} \right\|_\text{F}} { \left\| \m{P}_\m{V} \right\|_\text{F}}= \sqrt{2 - \frac{2\|\m{U}^\top\m{V}\|^2_\text{F}}{K}},
\end{align}
which is bounded in the range $\left[0,\sqrt{2}\right]$ with $\text{Err}(\m{U},\m{V})=0$ when the subspaces are identical and $\text{Err}(\m{U},\m{V})=\sqrt{2}$ when they are orthogonal.  We note that this notion of subspace error is a natural extension of the traditional distance between subspaces (see, e.g., Golub and Van Loan \cite[Section 2.6.3]{golub}) but is more quickly computed due to the use of Frobenius norm instead of operator norm.

For synthetic data generated as in Section \ref{sec:syn}, we have two senses of subspace error that we can measure.  First, given a finite set of data $\{\v{x}_t\}_{t=1}^N$, we can compute the top $K$ batch principal components $\m{U}^{(\text{batch})}_K$, which are the eigenvectors corresponding to the top $K$ eigenvalues of the batch covariance matrix
$$
\m{C}_\text{batch} = \frac{1}{N} \sum_{t=1}^N \v{x}_t\v{x}_t^\top.
$$
With these, we define the \emph{batch error}
\begin{align}\label{eq:errbatch}
\text{Err}_\text{batch}\left(\m{\widehat U}_K\right) &\equiv \text{Err}\left(\m{\widehat U}_K,\;\m{U}^{(\text{batch})}_K\right).
\end{align}
Additionally, in the case of synthetic data we also know the population principal components $\m{U}_*$, which are the eigenvectors corresponding to the top $K$ eigenvectors of $\m{C}$ in \eqref{eq:truecov}.  With these we define the \emph{population error}
\begin{align}\label{eq:errpop}
\text{Err}_\text{pop}\left(\m{\widehat U}_K\right) &\equiv \text{Err}\left(\m{\widehat U}_K,\;\m{U}_*\right).
\end{align}
For real data, the distinction above is less meaningful and so we consider only the batch error.

In all cases, we note that reported errors are obtained by orthogonalizing the columns of the estimate $\m{\widehat U}_K$ prior to computing the error.  In practice, the estimated principal components quickly become very close to orthogonal and so we do not observe a large change in value, but this orthogonalization step is technically necessary for our error measure to be a true measure of subspace error.

\section{Results}\label{sec:results}

All timing results were measured using Python 3.6 from the Anaconda distribution\footnote{\normalsize\url{https://anaconda.org/}} on a CentOS Linux workstation with a 12 core Xeon E5-2670 v3 2.30 GHz processor, 128GB of RAM.

To standardize the data in all cases, we center by subtracting the mean image (computed offline) and normalize by dividing by the mean norm of a centered data point (again offline), i.e., we compute
$$
\v{\bar x} = \frac{1}{N}\sum_{t=1}^N \v{x}_t \quad \text{and} \quad \nu = \frac{1}{N}\sum_{t=1}^N \|\v{x}_t-\v{\bar x}\|
$$
and standardize the data according to
$$
\v{x}_t \gets \frac{\v{x}_t - \v{\bar x}}{\nu}.
$$
While this standardization is not strictly necessary (and could instead be performed by varying the initial guesses and learning rates of the various algorithms), we find that in practice it is more convenient to fix the scale of the data for these experiments (though of course $\v{\bar x}$ and $\nu$ would have to be estimated incrementally in a truly online setting).

\subsection{Synthetic data}
\subsubsection{Simulation parameters}
For our synthetic data examples we use the spiked covariance model of Section \ref{sec:syn}.  For a test with $\m{U}_*\in\R^{D\times K}$, we choose the diagonal of $\m{S}$ as
\begin{align*}
S_{kk} = 1 - \frac{k-1}{2(K-1)}, \quad k=1,2,\dots,K,
\end{align*}
such that the largest ``clean'' entry is $S_{11}=1$ and the smallest is $S_{KK} = 1/2$.

To initialize the algorithms, we use the following.  Given a data stream $\{\v{x}_t\}$, we take the first $K$ data points and orthogonalize them to obtain the thin QR factorization
\begin{align}
\left[\begin{array}{c|c|c|c} \v{x}_1&\v{x}_2& \dots&\v{x}_K\end{array}\right] = \m{Q}\m{R}
\end{align}
where $\m{Q}\in\R^{D\times K}$ has orthonormal columns.  For IPCA and CCIPCA, we take the initial estimate of the principal components to be $\m{\widehat U}_K = \m{Q}$.  For IPCA the initial squared singular value estimate is taken as $\m{\widehat \Sigma}_K^2 = \m{0}$, and for CCIPCA we take $\hat\sigma_k^2 = 10^{-8}$ for each $k$, which were experimentally determined to be effective choices.  Because SM and FSM do not directly take either singular vector estimates or singular value estimates as input, the initialization for these algorithms is slightly different.  For SM, we chose $\m{M} = \frac{1}{100}\m{I}$ and $\m{W} = \frac{1}{100}\m{Q}^\top$ at initialization, such that each initial parameter is relatively small but $\m{M}^{-1}\m{W}=\m{Q}^\top$.  For FSM, we analogously chose $\m{M}_\text{inv}=100\m{I}$ and $\m{W}=\frac{1}{100}\m{Q}^\top$, which is equivalent to the SM case in exact arithmetic.

The algorithms all have learning rates to set.  For IPCA, we use the standard choice $\alpha_{t,\text{IPCA}} = 1/t$.  For CCIPCA we take the authors' suggested learning rate of
$$
\alpha_{t,\text{CCIPCA}} = \frac{1+\ell}{t},
$$
with ``amnesiac parameter'' $\ell=2$ \cite{ccipca}, though for the first few iterations where $t<1+\ell$ we modify this to be in the range $(0,1)$.  The algorithms SM and FSM take two learning rates which theoretically may be taken to be distinct \cite{neural_computation}.  In practice, we take equal learning rates
\begin{align}\label{eq:stepsize}
    \alpha_t = \beta_t = \frac{2}{\gamma t+5},
\end{align}
where there is sensitivity to the hyper-parameter $\gamma$.

As a centralized reference, the parameters that we vary in these examples are
\begin{itemize}
\item $D$: dimensionality of the data, i.e., $\v{x}_t\in\R^D$
\item $K$: dimensionality of the generated principal subspace, i.e., $\m{U}_*~\in~\R^{D\times K}$
\item $N$: number of data points generated (and thus number of iterations), i.e., $\{\v{x}_t\}_{t=1}^N$
\item $\rho$:  noise level as in \eqref{eq:truecov}
\item $\gamma$: learning rate hyper-parameter in \eqref{eq:stepsize}
\end{itemize}

\subsubsection{Timing}
First, we investigate the run   time performance of the different algorithms on these synthetic data models.  To do this accurately, we estimate the wall-clock time per iteration for single-threaded execution of each algorithm by averaging across a fixed number of iterations $N = 100$.  It should be noted that the accuracy achieved by the different algorithms is not identical with a fixed number of iterations, which means that this metric is not appropriate as a measure of time-to-solution for a fixed desired accuracy of the solution.  However, in the streaming data setting this time-per-iteration metric is crucial as it ultimately dictates bandwidth.

\begin{figure}
\centering
\includegraphics[width=8.75cm]{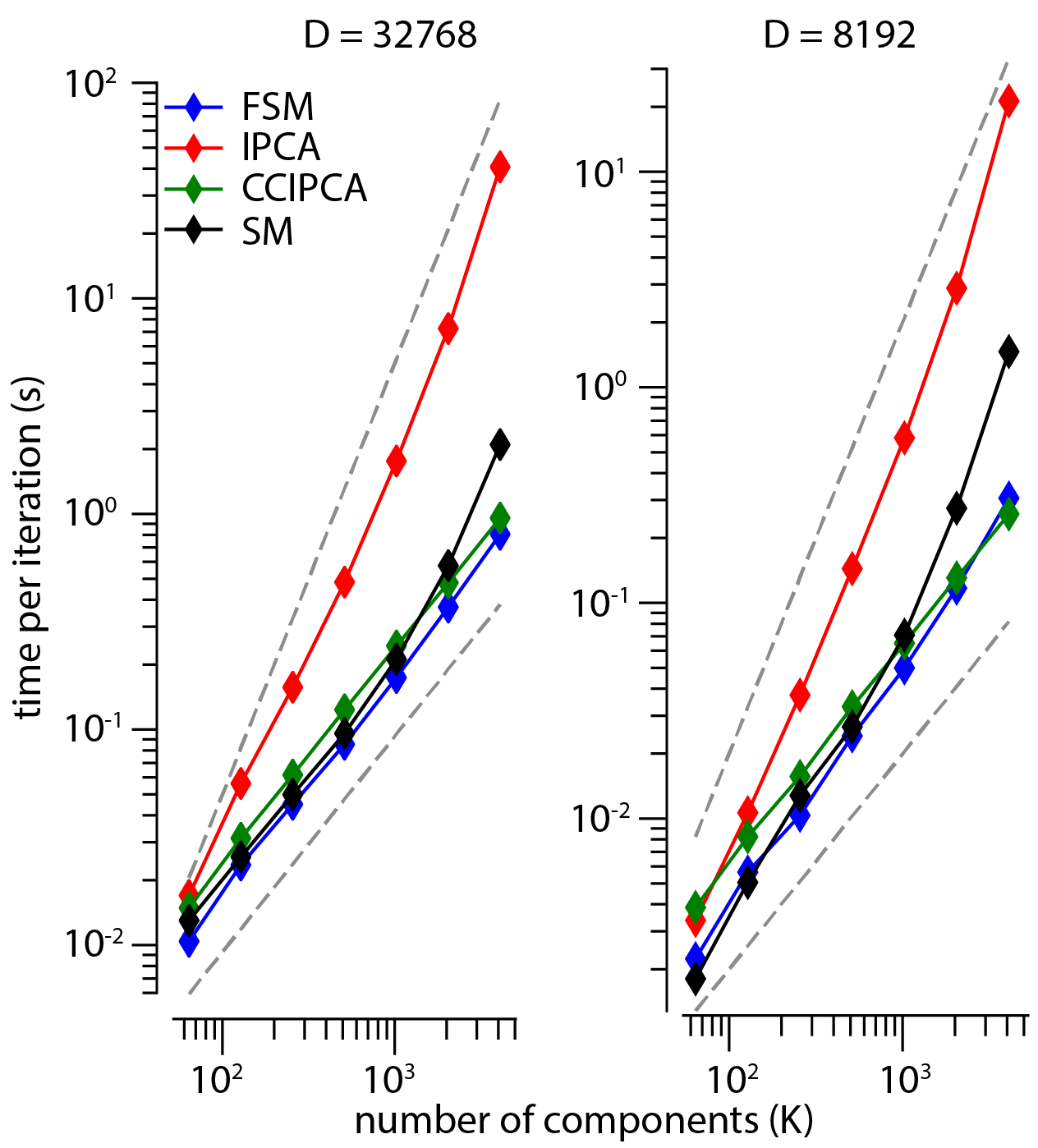}
\caption{Time per iteration of the different online algorithms in function of the input and output dimensionalities. Each data point represents the average time per iteration over 10 runs of 10 samples each, for a total of 100 averaged samples. The top trend line is $\O(K^2)$ and the bottom trend line is $\O(K)$.}
\label{fig:time_per_iter}
\end{figure}

In Figure \ref{fig:time_per_iter} we show the time-per-iteration results as we vary $K$ between $K=64$ and $K=4096$, where the left subplot shows the results for $D=32768$ and the right subplot shows the results for $D=8192$.  We take the noise level $\rho = 2\times 10^{-3}$, and for SM and FSM fix the learning rate parameter $\gamma=2$, though of course the runtime is independent of these choices.

We observe first that the empirical complexity of IPCA roughly matches its theoretical complexity of $\O(DK^2 + K^3)$: the runtime roughly follows the $\O(K^2)$ trend line until $K$ is large, at which point the growth exceeds $\O(K^2)$ in both plots.  In contrast, the complexity of CCIPCA closely follows the $\O(K)$ trend line, making CCIPCA much more efficient for large problems.  For SM with explicit inversion, we find that the initial runtime complexity is roughly $\O(K)$ like CCIPCA but for larger $K$ the complexity is greater and matches more closely that of IPCA.  This is to be expected, given the $\O(DK + K^3)$ theoretical complexity.  However, by incorporating the Sherman-Morrison formula to remove matrix inversion, we see that the runtime drops markedly for large $K$, giving FSM the best time-per-iteration on most examples with an empirical scaling of roughly $\O(K)$ as expected.

\subsubsection{Accuracy}
Having demonstrated the runtime differences in the various algorithms, we turn to accuracy, as measured by the subspace errors defined in Section \ref{sec:accuracy}.

\begin{figure}
\centering
\includegraphics[width=8cm]{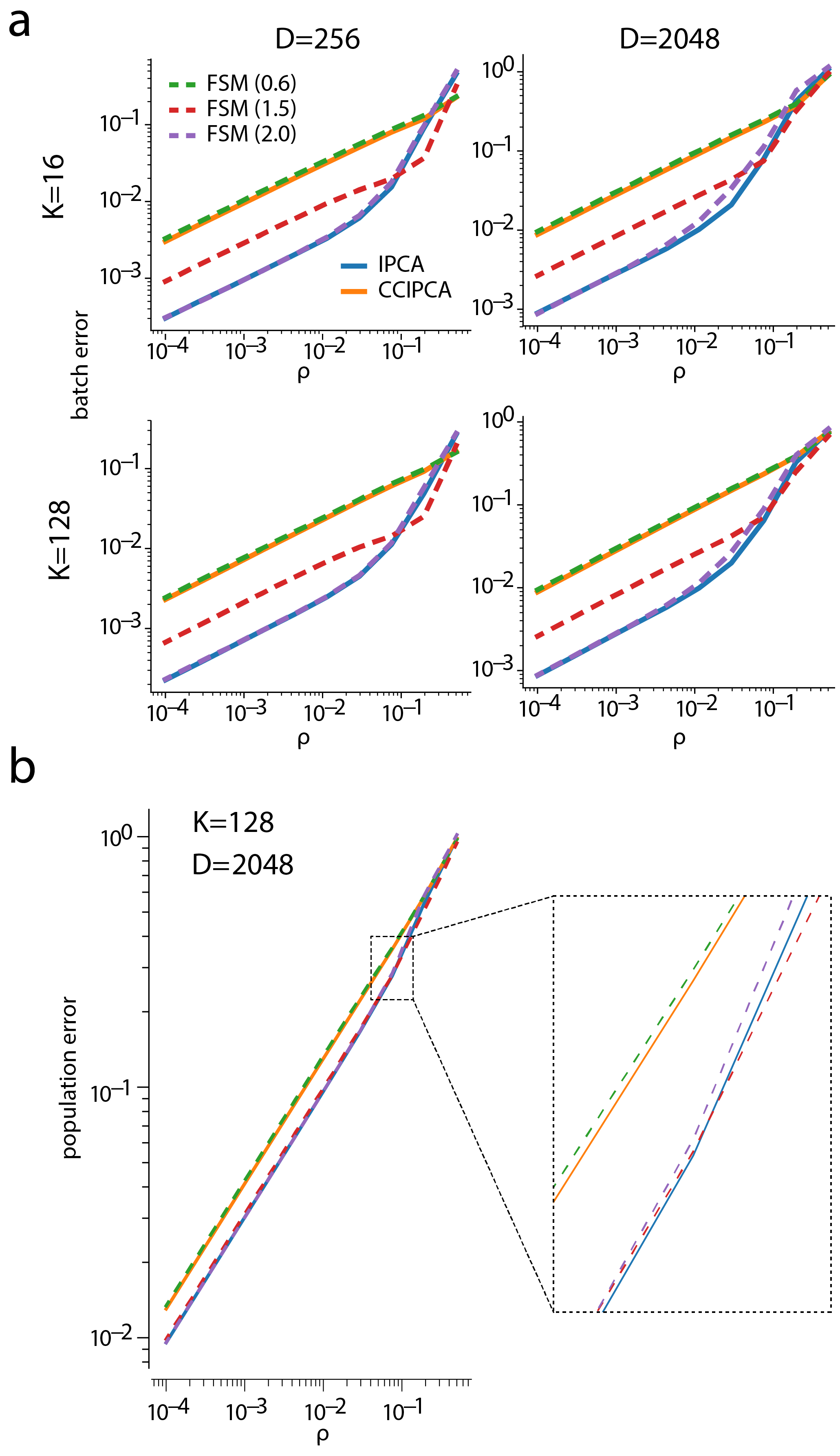}
\caption{Subspace error results compared to the batch (a) and population (b) principal components for the spiked covariance model with a fixed number of samples $N=6000$.  The three different curves for FSM correspond to three different choices of the learning rate hyper-parameter $\gamma$. \label{fig:spiked_err}}
\end{figure}

In Figure \ref{fig:spiked_err}a, we show $\text{Err}_\text{batch}$ in \eqref{eq:errbatch} after $N=6000$ iterations as a function of the noise level $\rho$ for each algorithm with a few combinations of $D$ and $K$.  For each curve, we average over 10 trials.  We omit results for SM and show only those for FSM, since they are identical.  However, because FSM has a free hyper-parameter $\gamma$ controlling the learning rate we show three different curves for $\gamma=0.6$ (green), $\gamma=1.5$ (red), and $\gamma=2.0$ (purple).   Intriguingly, we see that for different choices of $\gamma$ FSM performs the same as either of the other two algorithms, with $\gamma=0.6$ matching the performance of CCIPCA (yellow) and $\gamma=2.0$ matching that of IPCA (blue).  For this example, the best results are given by IPCA/FSM(2.0), beating CCIPCA/FSM(0.6) by about an order of magnitude except in the very noisy regime.

In Figure \ref{fig:spiked_err}b we show some corresponding $\text{Err}_\text{pop}$ results as defined in \eqref{eq:errpop}, using the same parameters as for the $\text{Err}_\text{batch}$ results of Figure \ref{fig:spiked_err}.  While we observe small variations in performance by this metric, the algorithms behave more-or-less equivalently with CCIPCA/FSM(0.6) behaving slightly worse than the other choices. This lends credence to our parameter choices.  We omit figures for the other combinations of $K$ and $D$, as they give effectively the same results.

\subsection{Real data}
For real-world datasets, we investigate the accuracy from both a quantitative and a qualitative perspective.  As remarked in Section \ref{sec:accuracy}, for real data we consider only the batch subspace error $\text{Err}_\text{batch}$. For these data, the primary parameters we consider are the number of components, $K$, and the step size hyper-parameter $\gamma$ for SM and FSM as defined in \eqref{eq:stepsize}.  We initialize the same as in the synthetic data examples.

\subsubsection{Accuracy, quantitative}\label{sec:realdata_acc_quant}
\begin{figure*}
\centering
\includegraphics[width=13.5cm]{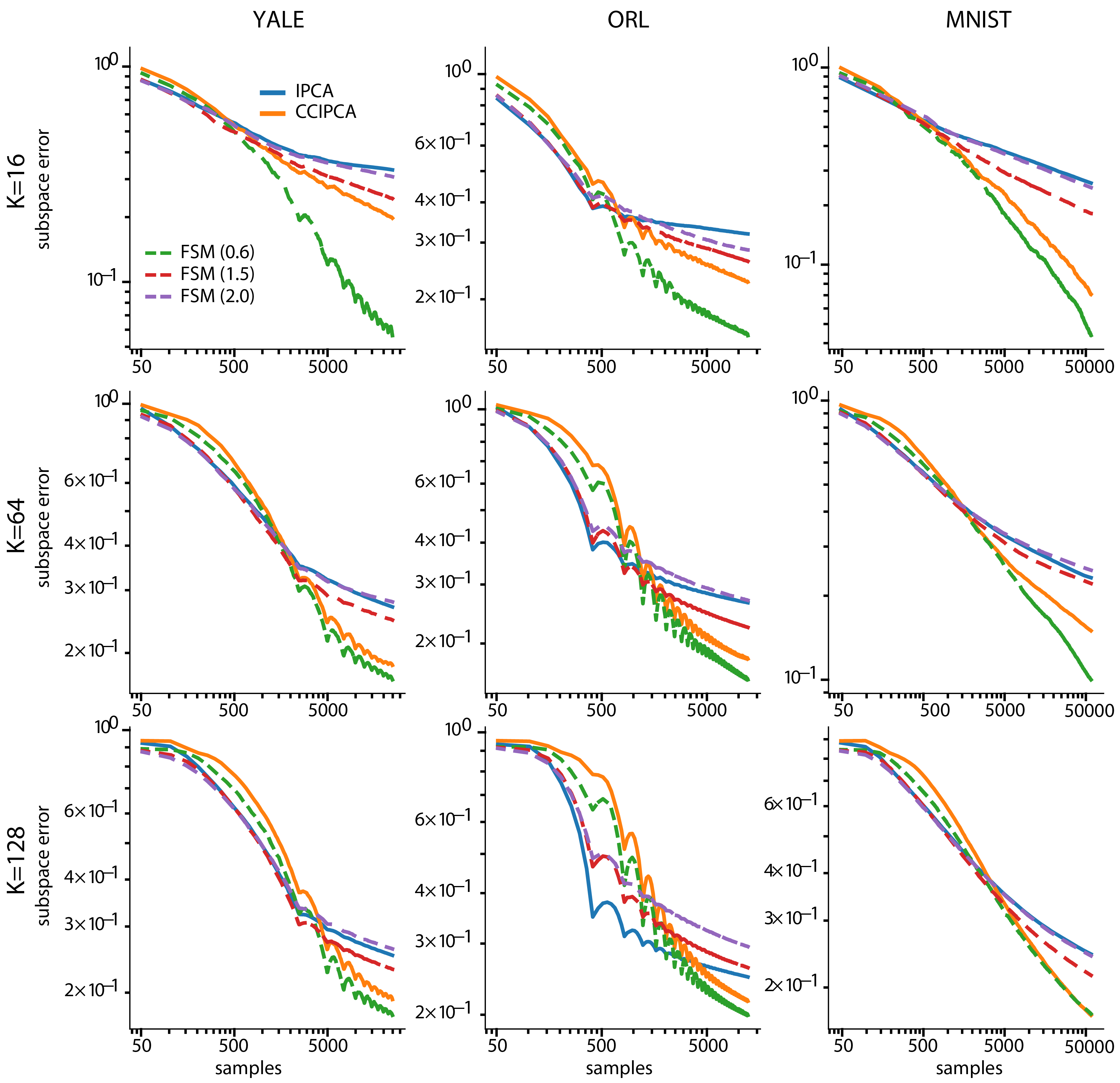}
\caption{For the different algorithms we show quantitative results on the YALE, ORL, and MNIST datasets. For each algorithm, we run ten times on random permutations of the datasets and take the median of the result.  The horizontal axis shows the sample index $t$ whereas the vertical axis shows the subspace error $\text{Err}_\text{batch}$.  We note that each subplot uses a different vertical axis.  The number of epochs used for YALE is 10, the number of epochs for ORL is 30, and for MNIST we use only one pass over the data.}
\label{fig:trajectories}
\end{figure*}

To demonstrate the performance of our algorithms, we show in Figure \ref{fig:trajectories} the ``trajectories'' $\text{Err}_\text{batch}$ as a function of the iteration index $t$, i.e., the number of samples seen.  Each shown curve is the median across 10 trials.  As before, we use three different choices for the learning rate hyper-parameter $\gamma$ for FSM.

Varying the number of components $K$ that we fit across a few different values, we see that in general the algorithms that give the best performance asymptotically are CCIPCA and FSM(0.6) and those that perform the worst are IPCA and FSM(2.0), in contrast to the synthetic data results.  Overall, for all algorithms the convergence is much slower than observed in the synthetic data examples, likely due to the lack of a pronounced spectral gap (in contrast to the spiked covariance model).

There is a discrepancy between algorithms that perform well with a small number of samples and those that perform well with a larger number of samples, as can be seen across the three different datasets.  In general, the trajectories for CCIPCA and FSM(0.6) initially lag those for the other algorithms, but ultimately FSM(0.6) gives the best performance with CCIPCA close behind.  The worst performers here are IPCA and FSM(2.0), which continue to converge steadily but at a slower rate.  Focusing just on the three FSM variants, we see a clear dependence on the learning rate hyper-parameter $\gamma$: if appropriately chosen, FSM performs as well as or better than CCIPCA, but if chosen incorrectly FSM performs as poorly as or worse than IPCA.  However, we note that the best choice of $\gamma$ is the same across all data sets, implying that $\gamma=0.6$ may be a robust choice under our normalization scheme.

\subsubsection{Accuracy, qualitative}
Because the order of magnitude of the subspace errors shown in Figure \ref{fig:trajectories} is relatively large, we additionally opt to perform a qualitative exploration of the accuracy to ensure that we are in a regime where these algorithms are useful.  To do this, we visualize the reconstruction of each image obtained by orthogonally projecting the image onto the current subspace estimate for each algorithm.

For very large projection errors of $\text{Err}_\text{batch} \gtrsim 0.75$, we observe qualitatively poor reconstruction results.  However, the reconstruction quality improves substantially with more samples, and we see that even a subspace error of $\sim 0.2$ gives reconstruction results that are visually comparable to the offline batch PCA results on the full dataset.  While the PCA accuracy required will differ depending on application, we view these qualitative results as a sanity check that the algorithms are giving reasonable subspaces.

\begin{figure*}
\centering
\includegraphics[width=14.5cm]{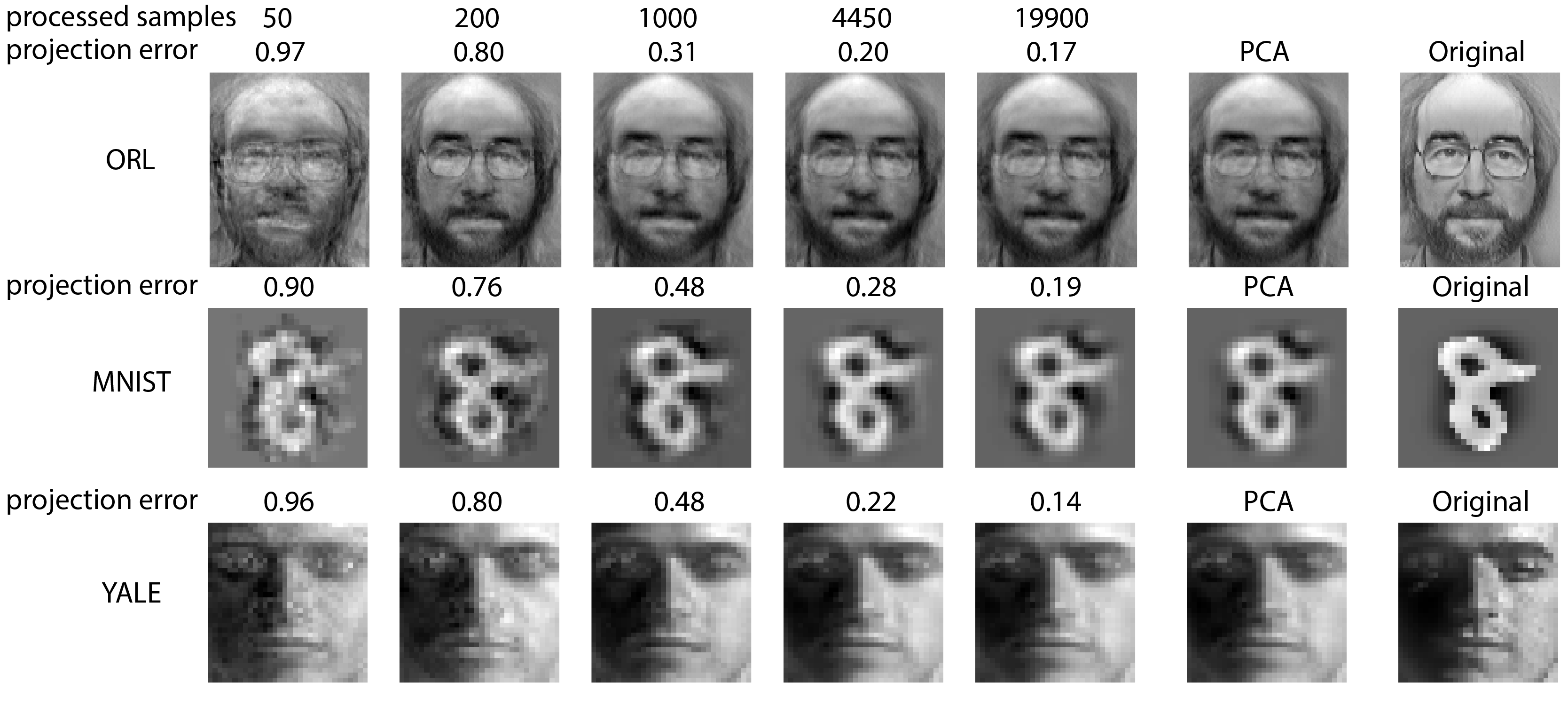}
\caption{Example of the evolution of the projection error and the corresponding reconstructions as a function of the number of processed samples for the ORL ($N=400$ images), YALE ($N=2400$) and MNIST ($N=1200$) datasets. The number of components was set to $K=64$. The images show how using the principal subspace at different points of training one can reconstruct a demeaned example image.}
\label{fig:evol_pe_mnist}
\end{figure*}

\section{Conclusions}
For PSP of big data in a streaming setting where data are presented one at a time, we have demonstrated numerically that SM can be a competitive algorithm both with synthetic data examples and real data sets.  To obtain optimal per-iteration time complexity, we introduced a variation of SM, FSM, which is numerically equivalent to SM but uses the Sherman-Morrison formula to substantial advantage.  A weakness of our approach is the lack of proof of convergence, though this is shared by IPCA.  Empirically, however, we find that both FSM/SM and IPCA converge in most of our tests, though the rate of convergence can be impacted by improper choice of learning rate.

From a computational standpoint, we observe that FSM outperforms IPCA by a large margin in terms of runtime while retaining accuracy.  Compared to CCIPCA, which has the same per-iteration complexity, FSM  with the appropriate parameter settings yields better performance in terms of subspace error.  Further, To reproduce the results of this paper and as a general tool for online PSP, we introduced the companion software package {\ttbs{online\_psp}}, which offers performant Python 3 implementations of the algorithms considered as well as a standardized testing framework. We provide a similarly performant version of the FSM, CCIPCA, and IPCA algorithms in the {\ttbs{online\_psp\_matlab}} MATLAB package.

In future work, we aim to better understand when convergence of SM and FSM can be assured and how to best choose the learning rates $\alpha_t$ and $\beta_t$ for optimal performance.

\section*{Acknowledgments}
The authors thank Mariano Tepper and Eftychios Pnevmatikakis for helpful input that contributed to the quality of this paper.

\bibliographystyle{IEEEtran}
\bibliography{pca_code}

\end{document}